\documentclass[10pt,aps,prl,twocolumn,superscriptaddress,amsmath,amssymb,showpacs]{revtex4-1}
\usepackage{bm}
\usepackage{amsmath}
\usepackage[dvips]{graphicx}
\begin{document}

\title{Quasiclassical approach to high-energy QED processes in strong laser and atomic fields}

\author{A. Di Piazza}
\email{dipiazza@mpi-hd.mpg.de}
\affiliation{Max-Planck-Institut f\"ur Kernphysik, Saupfercheckweg 1, 69117 Heidelberg, Germany}

\author{A. I. Milstein}
\email{milstein@inp.nsk.su}
\affiliation{Max-Planck-Institut f\"ur Kernphysik, Saupfercheckweg 1, 69117 Heidelberg, Germany}
\affiliation{Budker Institute of Nuclear Physics of SB RAS, 630090 Novosibirsk, Russia}

\date{\today}

\begin{abstract}
An approach, based on the use of the quasiclassical Green's function, is developed for investigating high-energy quantum electrodynamical processes in combined strong laser and atomic fields. Employing an operator technique, we derive the Green's function of the Dirac equation in an arbitrary plane wave and a localized potential. Then, we calculate the total cross section of high-energy electron-positron photoproduction in an atomic field of arbitrary charge number (Bethe-Heitler process) in the presence of a strong laser field. It is shown that the laser field substantially modifies the cross section at already available incoming photon energies and laser parameters. This makes it feasible to observe the analogous effect in a laser field of the Landau-Pomeranchuk-Migdal effect for the Bethe-Heitler process.
\end{abstract}

\pacs{12.20.Ds, 31.30.J-, 42.50.Xa}
\maketitle

The high intensity of available laser systems requires the investigation of the influence of laser light on fundamental QED processes in an atomic field like electron-positron ($e^+e^-$) photoproduction, the so-called Bethe-Heitler (BH) process, and bremsstrahlung \cite{Di_Piazza_2012}. This gives perspectives of testing QED in the presence of such intense fields, that they have to be taken into account beyond of perturbation theory. The cross section of QED processes in a pure atomic field depends on the energy of the incoming particle and on the parameter $Z\alpha$, where $Z$ is the atomic charge number and $\alpha\approx 1/137$ is the fine-structure constant (units with $\hbar=c=1$ are employed throughout). In general, the influence of the laser field is mainly characterized by the two parameters $\xi=|e|E/m\omega_0$ and $\chi=(\epsilon/m)(E/E_c)$. Here, $e$ and $m$ are the electron charge and mass, respectively, $E$ and $\omega_0$ are the laser's electric field amplitude and photon energy, respectively, $\epsilon$ is the energy of the incoming particle (a photon in the case of the BH process and an electron in the case of bremsstrahlung) and $E_c=m^2/|e|=1.3\times 10^{16}\;\text{V/cm}$ is the critical electric field of QED.

The influence of a laser field on QED processes occurring in an atomic field has already been widely studied in the literature, by including, however, only the laser field exactly (as a plane-wave), whereas the atomic field has been taken into account in the leading approximation in the parameter $Z\alpha$ (Born approximation). For instance, the BH process was considered in \cite{Loetstedt_2008,Di_Piazza_2009}, bremsstrahlung was investigated in \cite{Loetstedt_2007} (see also \cite{Karapetian_1978}, the book in \cite{Fedorov_b_1997} and references therein) and Delbr\"{u}ck scattering in \cite{Di_Piazza_2008}. Since all above-cited results have been obtained in the Born approximation, only the electron states exact in the plane-wave field (Volkov states) have been employed \cite{Landau_b_4_1982}. 

When the Dirac equation in an external field can be solved analytically (as in a constant field, in a plane wave, and in a pure Coulomb field), it is possible to obtain the amplitudes of a QED process via the operator technique, which does not imply the use of the explicit form of the electron's wave functions in the field. This method was developed in \cite{Schwinger_1951,Baier_1974,Baier_1975} in the case of a constant homogeneous electromagnetic field, in \cite{Baier_1975_b,Baier_1975_c} in the case of a plane wave, and in \cite{Milstein_1982} in the case of a Coulomb field. When the Dirac equation cannot be solved analytically, the amplitude of QED processes at high-energies can be obtained via the operator technique based on the applicability of the quasiclassical approximation \cite{Baier_b_1998}. An alternative method of calculation employs the quasiclassical Green's function of the Dirac equation in the external field (see \cite{Milstein_1983} for the case of a pure Coulomb field, \cite{LM95A} for an arbitrary spherically symmetric field, and \cite{LMS00} for a localized field which generally possesses no spherical symmetry).

In the present Letter we generalize the method of the quasiclassical Green's function to the case of QED processes occurring in a localized field and in a plane wave. We derive the quasiclassical Green's function of the Dirac equation, describing the propagation of an ultrarelativistic electron, almost counterpropagating with respect to the plane wave. This configuration is the most relevant from an experimental point of view. We emphasize that both the localized and plane-wave fields are included exactly in the calculations, which are performed in the leading-order with respect to the parameter $m/\epsilon\ll 1$. The obtained Green's function is then applied to calculate the total cross section of the BH process exactly in the parameters of the atomic and the laser field. We study the problem in the rest frame of the atom assuming that the energy $\omega$ of the incoming photon largely exceeds $m$, that $\omega_0\omega\ll m^2$ (i.e., there is no pair production stemming from one laser photon and the incoming photon), and that both $Z\alpha$ and $\chi=(\omega/m)(E/E_c)$ are of the order of unity. In this situation the total cross section of the pair-production process becomes independent of the laser parameter $\xi$ and, since $E/E_c\ll 1$, the probability of pair production only from laser and atomic field \cite{Yakovlev_1965} is exponentially suppressed. Also, we split the total probability of the process into two terms, the first being the total probability of $e^+e^-$ pair photoproduction in the laser field (at $Z\alpha=0$), and the second representing the influence of the laser field on the BH process. Only the latter term is considered here, as the first process has already been investigated in detail \cite{Reiss_1962,Nikishov_1964}. We show that the presence of the laser field substantially modifies the cross section of BH process even at moderate values of the parameter $\chi$, corresponding to already available laser intensities and photon energies. The considered process indicates the feasibility of successfully applying the obtained Green's function for investigating a wide class of QED problems, including processes in the combined field of an intense laser and a highly-charged ion.
 
By considering for definiteness a plane wave propagating in the direction antiparallel to the $z$-axis, it is convenient to pass from the variables $t$ and $z$ to the variables $\phi=t-z$ and $T=(t+z)/2$. In this way,
\begin{eqnarray}
&& p^0=i\partial_t=-p_\phi-p_T/2\, ,\;\;
p^z=-i\partial_z=-p_\phi+p_T/2\,,\nonumber\\
&&p_\phi=-i\partial_\phi\,,\quad p_T=-i\partial_T\,, \quad \bm p_\perp=-i\partial_{\bm\rho}\,,
\end{eqnarray}
where $\bm\rho$ is the component of the vector $\bm r$ perpendicular to $\bm{z}$. We first analyze the propagation of an ultrarelativistic charged scalar particle in a localized potential $V(\bm r)=V(\bm\rho,\, z)$ and in a plane-wave field described by the vector potential $\bm A(t+z)=\bm A(2T)$. We assume that the momentum of the particle is almost parallel to $\bm{z}$. In this case, we can replace the coordinate $z$ in the argument of the potential by $T$. The propagation of the particle is described by the Green's function of the Klein-Gordon equation,
\begin{eqnarray}\label{KGV}
&&D^{(0)}(x,x')=\int\frac{d\epsilon}{2\pi}e^{-i\epsilon(\phi-\phi')} D^{(0)}(T,\,\bm\rho|\, T',\,\bm\rho';\,\epsilon)\,,\nonumber\\
&&D^{(0)}(T,\,\bm\rho|\, T',\,\bm\rho';\,\epsilon)=
\frac{1}{{\cal P}^2-m^2+i0}\delta(T-T')\delta(\bm\rho-\bm\rho')\nonumber\\
&&=-i\int_0^\infty ds\,\exp\left[is({\cal P}^2-m^2)\right]\delta(T-T')\delta(\bm\rho-\bm\rho')\,,\nonumber\\
&&\equiv-i\int_0^\infty ds\,e^{-ism^2}F(T,\,\bm\rho|\, T',\,\bm\rho')\,.\nonumber\\
&&{\cal P}^2=-2\epsilon V(\bm\rho,\,T)-H,\;\; H=2\epsilon p_T+[\bm p_\perp-\bm{\mathcal{A}}(T)]^2\,,
\end{eqnarray}
where $\bm{\mathcal{A}}(T)=e\bm A(2T)$. By neglecting the term $V^2(\bm\rho,\,T)$ in comparison with $\epsilon V(\bm\rho,\,T)$ and the term $\{p_T,V\}=p_T\,V+V\,p_T$ in comparison with $\epsilon p_T$ in Eq. (\ref{KGV}), we represent the function $F(T,\,\bm\rho|\, T',\,\bm\rho')$ as
\begin{eqnarray}\label{F}
&& F(T,\,\bm\rho|\, T',\,\bm\rho')=L(s)\,e^{-isH}\delta(T-T')\delta(\bm\rho-\bm\rho')\,,\nonumber\\
&&L(s)=e^{-is[2\epsilon V(\bm\rho,\,T)+H]}e^{isH}\,.
\end{eqnarray}
The operator $L(s)$ satisfies the equation
\begin{align}
\frac{dL(s)}{ds}=&-2i\epsilon L(s)V(\bm\zeta_1,\,T-2\epsilon s)\,,\nonumber\\
\bm\zeta_1=&\bm\rho-2 s\bm p_\perp+2s\int_0^1dx\bm{\mathcal{A}}(T-2s\epsilon x)\,.
\end{align}
In the leading quasiclassical approximation, we can ignore the non-commutativity of the operators $V(\bm\zeta_1,\,T-2\epsilon s)$ at different values of $s$. As a result, we have
\begin{eqnarray}\label{LL1}
&&L(s)=\exp\left[-2is\epsilon\int_0^1dx V(\bm\zeta_2,\,T-2\epsilon sx)\right]\,,\nonumber\\
&&\bm\zeta_2=\bm\rho-2 sx\bm p_\perp+2sx\int_0^1dx'\bm{\mathcal{A}}(T-2s\epsilon xx')\,.
\end{eqnarray}
By employing the relation (see \cite{BKMS75})
\begin{equation}
e^{-isH}=\exp\Big\{-is\int_0^1dx[\bm p_\perp-\bm{\mathcal{A}}(T-2\epsilon s x)]^2 \,\Big\}\,e^{-2is\epsilon p_T},
\end{equation}
we obtain
\begin{eqnarray}\label{rel1}
&&e^{-isH}\delta(T-T')\delta(\bm\rho-\bm\rho')=-\frac{i}{4\pi s}\delta(T-T'-2\epsilon s)
\nonumber\\
&&\times\exp\left[i\frac{(\bm\rho-\bm\rho')^2}{4s}+is(\bm f^2-g^2)+i(\bm\rho-\bm\rho')\cdot\bm f\right]\,,\nonumber\\
&&\bm f=\int_0^1dx\,\bm{\mathcal{A}}(\tau)\,,\quad g^2=\int_0^1dx\,\bm{\mathcal{A}}^2(\tau)\,,
\end{eqnarray}
where $\tau=T-(T-T')x$. Then, we substitute Eqs. (\ref{LL1}) and (\ref{rel1}) in Eq. (\ref{F}) and we find
\begin{eqnarray}\label{F1}
&& F(T,\,\bm\rho|\, T',\,\bm\rho')=-\frac{i}{4\pi s}\delta(T-T'-2\epsilon s)\nonumber\\
&&\times\exp\Big[i\frac{(\bm\rho-\bm\rho')^2}{4s}+is(\bm f^2-g^2)+i(\bm\rho-\bm\rho')\cdot\bm f\Big]\,,\nonumber\\
&&\times\exp\left[-2is\epsilon\int_0^1dx V(\bm\zeta_3,\,\tau)\right]\times 1\,,\nonumber\\
&&\bm\zeta_3=\bm\zeta_2-2sx\Big(\bm f+\frac{\bm\rho-\bm\rho'}{2s}\Big)\,,
\end{eqnarray}
where $1$ stands for a constant function with unit value. Analogously to the case considered in \cite{LMS00}, it is important to account for the operator $\bm p_\perp$ in the argument of the potential in Eq. (\ref{F1}) only for $x$ close to $x_0=T/(T-T')$, when $0<x_0<1$, otherwise the operator $\bm p_\perp$ can be omitted.
By exploiting this fact, we can write 
\begin{eqnarray}\label{F2}
&&\exp\left[-2is\epsilon\int_0^1dx V(\bm\zeta_3,\,\tau)\right]\nonumber\\
&&=\exp\left(-i\frac{x_0s}{1-x_0}\bm p_\perp^2\right)\exp\left[-2is\epsilon\int_0^1dx V(\bm\zeta_4,\,\tau)\right]\,,\nonumber\\
&&\bm\zeta_4=\bm\rho -x(\bm\rho-\bm\rho')+2sx(1-x)\tilde{\bm f}\,,\nonumber\\
&&\tilde{\bm f}=\int_0^1dy\,[\bm{\mathcal{A}}(Ty)-\bm{\mathcal{A}}(T'y)]\,.
\end{eqnarray}
The useful identity
\begin{equation}
 e^{-i\gamma \bm p_\perp^2} g(\bm\rho)=\int\frac{d^2q}{i\pi}e^{iq^2}g(\bm\rho+2\sqrt{\gamma}\bm q),
\end{equation}
valid for an arbitrary function $g(\bm \rho)$ and a positive constant $\gamma$, can be employed, with $\bm q$ being a vector in the plane perpendicular to the vector $\bm{z}$. In this way, the integral over $s$ in Eq. (\ref{KGV}) can be performed, and the final result for the Green's function $D_0(T,\,\bm\rho|\, T',\,\bm\rho';\,\epsilon)$ is
\begin{eqnarray}\label{D0finalL}
&&  D^{(0)}(T,\,\bm\rho|\, T',\,\bm\rho';\,\epsilon)=\frac{i\theta(s_0)}{4\pi^2|T-T'|}
\exp\Big[is_0(\bm f^2-g^2)\nonumber\\
&&+i(\bm\rho-\bm\rho')\cdot\bm f +i\frac{(\bm\rho-\bm\rho')^2}{4s_0}-im^2s_0\Big]\nonumber\\
&&\times\int d^2q\,e^{iq^2}\exp\left[-2is_0\epsilon\int_0^1dx V(\bm\zeta_x,\,\tau)\right]\,,\nonumber\\
&&\bm\zeta_x=\bm\rho-x(\bm\rho-\bm\rho')+2\sqrt{\beta}\,\bm q +2\beta \tilde{\bm f}\,,\nonumber\\
&&s_0=\frac{T-T'}{2\epsilon}\,,\,\, x_0=\frac{T}{T-T'}\,,\,\,\beta=x_0(1-x_0)s_0\,,
\end{eqnarray}
with the functions $\bm{f}$, $g^2$ and $\tilde{\bm{f}}$ being given in Eqs. (\ref{rel1}) and (\ref{F2}) and with $\theta(x)$ being the step function. We note that $D_0(T,\,\bm\rho|\, T',\,\bm\rho';\,\epsilon)$ does not contain anymore momentum operators and we recall that the terms proportional to $\bm q$ and $\tilde{\bm{f}}$ are to be omitted if $x_0<0$ or $x_0>1$. As it should be, the Green's function $D_0(T,\,\bm\rho|\, T',\,\bm\rho';\,\epsilon)$ is invariant under the replacement $T\leftrightarrow T'$, $\bm\rho\leftrightarrow \bm\rho'$, and $\bm{\mathcal{A}}\rightarrow -\bm{\mathcal{A}}$. Also, if $\bm{\mathcal{A}}=\bm{0}$, it is in agreement with the corresponding result obtained in the quasiclassical approximation in \cite{LM95A,LMS00}, whereas, if $V=0$, it reduces to the exact Green's function in a laser field \cite{Schwinger_1951}.

We consider an optical laser field ($\omega_0\sim 1$~eV) and a particle with energy $\epsilon$ such that  $\epsilon\omega_0\ll m^2$ (i.e., $\epsilon\ll 100\;\text{GeV}$). It follows from  Eq. (\ref{D0finalL}) that in this case the typical value of $s_0$ is $s_0\sim 1/m^2$, such that $|T-T'|\sim (\epsilon/m)\lambda_C$ and $|\bm\rho-\bm\rho'|\sim\lambda_C$, where $\lambda_C=1/m$ is the Compton wavelength. Besides, it is $\beta\sim 1/m^2$, $\beta|\tilde{\bm f}|\sim \beta\omega_0|T-T'|\, |\bm{\mathcal{A}}|\sim \chi\lambda_C$, and $s_0(\bm{f}^2-g^2)\sim \chi^2$, where $\chi=(\epsilon/m)(E/E_c)\lesssim 1$ such that the effect of the laser field may be important. At $|\bm\rho+\bm\rho'|\gg \lambda_C$, we can neglect the terms proportional to $\bm q$ and $\tilde{\bm f}$ in $\bm{\zeta}_x$ and perform the integral over $\bm q$. The result is nothing more than the Green's function in the eikonal approximation.

The Green's function of the Dirac equation can be represented as
\begin{equation}\label{FGD}
G(x,\,x')=\langle x|(\hat{\cal P}-m+i0)^{-1}|x'\rangle=(\hat{\cal P}+m)D(x,\,x').
\end{equation}
Here, $\hat{\cal P}=\gamma^\mu{\cal P}_\mu$, where $\gamma^\mu$ are the Dirac matrices, ${\cal P}_\mu=p_\mu-eA_\mu(x)$, with $eA^{\mu}=(V(\bm{\rho},T),\bm{\mathcal{A}}(T))$, and $D(x,\,x')=\langle x|(\hat{\cal P}^2-m+i0)^{-1}|x'\rangle$ is the Green's function of the ``squared'' Dirac equation. It has been shown in \cite{LM95A,LMS97} that it is convenient to calculate the amplitudes of various QED processes in terms of the function $D(x,\,x')$, and we derive here its explicit form for the field under consideration. We have 
\begin{equation}
{\hat{\cal P}}^2={\cal P}^2+i\bm\alpha\cdot \bm\nabla V(\bm \rho,\, T)+\frac{i}{2}\hat\varkappa\, 
\bm \gamma\cdot\partial_T\bm{\mathcal{A}}(T)\,,
\end{equation}
where $\bm\nabla=\partial_{\bm\rho}+\bm{z} \partial_T$, and $\hat\varkappa=\gamma^0+\gamma^3$. We introduce the effective potential $\tilde V=V+\delta V$, where
\begin{equation}
\delta V=-\frac{i}{2\epsilon}\left[\bm\alpha\cdot \bm\nabla V(\bm \rho,\, T)+\frac{1}{2}\hat\varkappa\, 
\bm \gamma\cdot\partial_T\bm{\mathcal{A}}(T)\right]\,.
\end{equation}
Although $\delta V$ is small in comparison with $V$, it has a different matrix structure and the linear terms in $\delta V$ have also to be taken into account in the calculations (see, e.g., \cite{LMS00}). In order to obtain the corresponding correction, we can employ Eq. (\ref{D0finalL}) with the replacement $V\rightarrow\tilde V$ and then perform the expansion with respect to $\delta V$. As a result we find
\begin{eqnarray}\label{DfinalL}
&&D(T,\,\bm\rho|\, T',\,\bm\rho';\,\epsilon)=\Big\{1-\frac{i}{2\epsilon}\bm\alpha\cdot(\partial_{\bm\rho}+
\partial_{\bm\rho'})\nonumber\\
&&-\frac{\hat\varkappa}{4\epsilon}\bm\gamma\cdot[\bm{\mathcal{A}}(T)-\bm{\mathcal{A}}(T')]\Big\}D^{(0)}(T,\,\bm\rho|\, T',\,\bm\rho';\,\epsilon)\,.
\end{eqnarray}
Note that this expression is exact with respect to the parameters of both the laser field and the atomic potential.

Now, we apply the Green's function in Eq. (\ref{DfinalL}) to calculate the total photoproduction cross section $\sigma$ in the laser and the atomic electric field. Due to the optical theorem and following Ref. \cite{LM95A}, we can write $\sigma$ as
\begin{eqnarray}\label{s2}
&& \sigma=\frac{\alpha}{\omega}\mbox{Re}\int_0^\omega d\epsilon \int dT\,dT'd\bm\rho\,d\bm\rho'\,\theta(T-T')\nonumber\\ 
&&\times\mbox{Sp}\Big\{[(2\bm e^*\cdot{\bm{\mathcal{P}}}_\perp-\hat e^*\hat k) \, D(T,\,\bm\rho|\, T',\,\bm\rho';\,\epsilon)]\nonumber\\
&&\times [(2\bm e\cdot{\bm{\mathcal{P}}}'_\perp+\hat e\hat k) \, D(T',\,\bm\rho'|\, T,\,\bm\rho;\,\epsilon-\omega)]\Big\}\,,\nonumber\\
&&{\bm{\mathcal{P}}}_\perp=  -i\partial_{\bm\rho}-{\bm{\mathcal{A}}}(T)\,,\,\,
{\bm{\mathcal{P}}}'_\perp=  -i\partial_{\bm\rho'}-{\bm{\mathcal{A}}}(T')\,.
\end{eqnarray}
where $e^{\mu}=(0,\bm e)$ and $k^{\mu}=(\omega,\bm{k})$ are the polarization four-vector and the four-momentum of the incoming photon, respectively, and where the $\theta$-function in Eq. (\ref{D0finalL}) is taken into account. Below, we consider the case of incoming unpolarized photons. By substituting the Green's function (\ref{DfinalL}) in Eq. (\ref{s2}) and by performing the trace, we obtain
\begin{equation}\label{sigma1}
\begin{split}
\sigma=&\frac{8\alpha}{\omega}\mbox{Re}\int_0^\omega d\epsilon\int dTdT'd\bm\rho\,d\bm\rho'\,
\theta(T-T')\\
&\times\Big\{[{\bm{\mathcal{P}}}_\perp D_+^{(0)}]\cdot[{\bm{\mathcal{P}}}'_\perp D_-^{(0)}]
-\frac{\omega^2}{4\epsilon(\omega-\epsilon)}\\
&\times [({\bm{\mathcal{P}}}_\perp-{\bm{\mathcal{P}}}^{'*}_\perp) D_+^{(0)}]\cdot[({\bm{\mathcal{P}}}_\perp'-{\bm{\mathcal{P}}}^{*}_\perp) D_-^{(0)}]\Big\}\,,
\end{split}
\end{equation}
where $D_+^{(0)}=D^{(0)}(T,\,\bm\rho|\, T',\,\bm\rho';\,\epsilon)]$ and $D_-^{(0)}=D^{(0)}(T',\,\bm\rho'|\, T,\,\bm\rho;\,\epsilon-\omega)]$. Since we investigate the influence of the laser field on the cross section of the BH process, we assume the subtraction from the integrand in Eq. (\ref{sigma1}) of its value at $\bm A=\bm{0}$. Then, we exploit the relation $\omega_0T\sim \omega_0\omega/m^2\ll 1$, and expand the vector $\bm{\mathcal{A}}(T)$ so that $\bm{\mathcal{A}}(T)=\bm{\mathcal{A}}(0)+\bm{\mathcal{E}}T$, where the quantity $\bm{\mathcal{E}}=\partial_T\bm{\mathcal{A}}$ at $T=0$ depends on some constant phase (note that $\bm{\mathcal{E}}=-2e\bm{E}$, where $\bm{E}$ is the electric field's amplitude of the laser wave). The final result should be averaged over this phase. Due to gauge invariance, the cross section $\sigma$ is independent of $\bm{\mathcal{A}}(0)$. Also, it depends on $\bm{\mathcal{E}}$ via its square value $\bm{\mathcal{E}}^2$, which in turn, for a circularly polarized laser field, is independent of the phase of the field. The integrations in Eq. (\ref{sigma1}) are performed in a similar way as described in detail in \cite{LM95A, LMS00}, and we present here only the final result. In order to render the parameter $\chi$ as large as possible, we consider the case where $\omega/m^2\gg r_c$, where $r_c$ is the atomic screening radius. In the Thomas-Fermi model, $r_c \sim (m\alpha)^{-1}Z^{–1/3}$ and the cross section $\sigma$ has the form,
\begin{eqnarray}\label{MDfinal}
&& \sigma=\sigma_0\left\{\Phi(\chi)\left[\ln(183 Z^{-1/3})-\frac{1}{42}-f(Z\alpha)\right]+\Psi(\chi)\right\}\,,\nonumber\\
&&\sigma_0=\frac{\alpha(Z\alpha)^2}{m^2}\,,\,\,f(Z\alpha)=\mbox{Re}\,\psi(1+iZ\alpha)+C\,,
\end{eqnarray}
where $\psi(t)=d\ln\Gamma(t)/dt$, $C=0.577\dots$ is the Euler constant, $\Psi(\chi)=\Psi_1(\chi)+\Psi_2(\chi)$ and 
\begin{widetext}
\begin{eqnarray}\label{MCfinal}
&&\Phi(\chi)=16\,\mbox{Im}\int_0^1 dx \int_0^\infty d\tau\, e^{-i\varphi}
\left[-\frac{1}{4}+\frac{1}{3}a+2i\chi^2\tau^3 a^2\left(1-\frac{19}{15}a\right)\
+\frac{32}{15}\chi^4\tau^6a^4(1-a)\right]\,,\, \varphi=\tau+\frac{4}{3}\chi^2a^2\tau^3\,\nonumber\\
&&\Psi_1(\chi) =\mbox{Im}\int_0^1 dx \int_0^\infty d\tau e^{-i\varphi}
\bigg\{8\left[-\log(\tau)-C-i\frac{\pi}{2}\right]\left[-\frac{1}{4}+\frac{1}{3}a+2i\chi^2\tau^3 a^2\left(1-\frac{19}{15}a\right)+\frac{32}{15}\chi^4\tau^6a^4(1-a)\right]\nonumber\\
&&-\frac{2}{21}+\frac{4a}{7}+i\frac{16}{3}\chi^2\tau^3a^2\left(\frac{8}{7}-\frac{461}{175}a\right)+
\frac{14464}{525}\chi^4\tau^6a^4(1-a)\bigg\}\,, \quad a=x(1-x)\,,\quad \chi=\frac{\omega|\bm{\mathcal{E}}|}{2m^3}=\frac{\omega}{m}\frac{E}{E_c}\,,\nonumber\\
&&\Psi_2(\chi) =\frac{4}{\pi}\mbox{Re}\int_0^1dy\int_0^1 dx \int_0^\infty d\tau e^{-i\varphi}
\!\!\!\int d^2 q \ln\bigg(\frac{|\bm{q}+\bm F|}{q}\bigg)\,e^{iq^2}\,(\bm q+\bm F)^2\,
\big[a (2y-1)^2q^2+(1-a)(\bm q+\bm F)^2-i\big].
\end{eqnarray}
\end{widetext}
The vector $\bm F$ in $\Psi_2(\chi)$ is equal to $4\chi a\sqrt{y(1-y)}\tau^{3/2}\bm s$, with $\bm s$ being an arbitrary unit vector, $\bm s^2=1$. Also, the integral over $\bm q$ in $\Psi_2(\chi)$ can be performed analytically, but we do not report the result here for the sake of brevity. We note that all the Coulomb corrections in $\sigma$ are enclosed in the function $f(Z\alpha)$, as in the case of a pure atomic field, and that the effects of the laser field are included in the functions $\Phi(\chi)$ and $\Psi(\chi)$. The asymptotic forms of these functions at $\chi\ll 1$ are
\begin{align}\label{as}
\Phi(\chi)=\frac{28}{9}\Big(1+\frac{2136}{1225}\chi^2\Big), && \Psi(\chi)=-\frac{312752}{55125}\chi^2.
\end{align}
At $\chi\gg 1$ it is $\Phi(\chi)\propto \chi^{-2/3}$ and $\Psi(\chi)\propto \chi^{-2/3}\ln\chi$ but these asymptotics are applicable only at very large values of $\chi$. Figure 1 displays the dependence of the ratio $\sigma(\chi)/\sigma(0)$ on $\chi$ for $Z=83$, where $\sigma(\chi)$ is given in Eq. (\ref{MDfinal}). We have seen that this ratio depends very weakly on $Z$. The figure indicates that high-order effects in $\chi$ become significant already at relatively small values of $\chi$. While the cross section results amplified for $\chi<0.5$ (up to about $10\%$ at $\chi=0.25$), it is suppressed by about $40\%$ already at $\chi=1$. This suppression is the analogous in the laser field of the Landau-Pomeranchuk-Migdal (LPM) effect due to multiple scattering of a charged particle in matter (see the Review in \cite{Baier_2005}). Unless the LPM effect for BH process in matter, which has never been observed because of the required ultra-high photon energies ($\omega\gtrsim 2.5\;\text{TeV}$), the effect in the laser field can be observed in principle at $\omega=10\;\text{GeV}$ for an already available intensity of $10^{21}\;\text{W/cm$^2$}$ \cite{Di_Piazza_2012}.
\begin{figure}
\begin{center}
\includegraphics[width=0.7\linewidth]{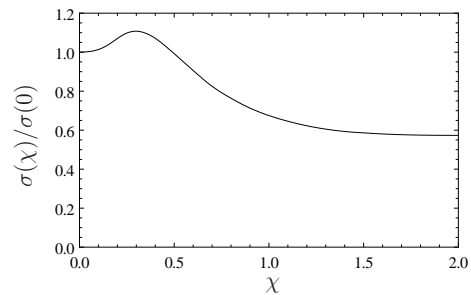}
\caption{The dependence of the ratio $\sigma(\chi)/\sigma(0)$ on the parameter $\chi$ for $Z=83$, where $\sigma(\chi)$ is given in Eq. (\ref{MDfinal}).}\label{Phi}
\end{center}
\end{figure}
In conclusion, we have derived the Green's function of the Dirac equation in the quasiclassical approximation by taking into account exactly the parameters of a strong plane wave and of a localized field. By employing this Green's function we have determined the probability of the BH process exactly in these parameters. The laser field is found to suppress the cross section by about $40\%$ already at $\chi\gtrsim 1$ and this renders in principle feasible the observation of the analogous of the LPM effect in a laser field for the BH process at already available energy of the incoming photon and parameters of the laser field. Finally, the considered example shows the feasibility of employing the Green's function in Eq. (\ref{DfinalL}) to investigate exactly in the parameters of both the laser and the localized field, such processes, which so far have been treated in the Born approximation in $Z\alpha$.

A. I. M. gratefully acknowledges the Max-Planck-Institute for Nuclear Physics for warm hospitality and financial support during his visit. The work was supported in part by  the Ministry of Education and Science of the Russian Federation and  the Grant 14.740.11.0082 of Federal special-purpose program “Scientific and scientific-pedagogical personnel of innovative Russia”.

\end{document}